\documentclass[%
 reprint,
superscriptaddress,
nofootinbib,
bibnotes,
 amsmath,amssymb,
 aps,
 prl,
]{revtex4-2}

\usepackage{graphicx}
\usepackage{dcolumn}
\usepackage{bm}
\usepackage[mathlines]{lineno}
\usepackage{verbatim} 
\usepackage{xcolor}
\usepackage{soul}

\DeclareUnicodeCharacter{2061}{}
\begin{document}

\title{Dense Suspension Inertial Microfluidic Particle Theory (DENSE-IMPACT) Model for Elucidating Outer Wall Focusing at High Cell Densities}

\author{Soon Wei Daniel Lim}
\affiliation{
 Department of Molecular and Cellular Physiology, Stanford University, Stanford, California, 94305, USA
}
\author{Yong How Kee}
\affiliation{
Bioprocessing Technology Institute (BTI), Agency for Science, Technology and Research, 20 Biopolis Way, Singapore 138668, Singapore
}

\author{Scott Nicholas Allan Smith}
\affiliation{
Bioprocessing Technology Institute (BTI), Agency for Science, Technology and Research, 20 Biopolis Way, Singapore 138668, Singapore
}

\author{Shan Mei Tan}
\affiliation{
Bioprocessing Technology Institute (BTI), Agency for Science, Technology and Research, 20 Biopolis Way, Singapore 138668, Singapore
}

\author{An Eng Lim}
\affiliation{
Bioprocessing Technology Institute (BTI), Agency for Science, Technology and Research, 20 Biopolis Way, Singapore 138668, Singapore
}

\author{Yuansheng Yang}
\affiliation{
Bioprocessing Technology Institute (BTI), Agency for Science, Technology and Research, 20 Biopolis Way, Singapore 138668, Singapore
}
    
\author{Shireen Goh}
\email{shireen\_goh@sutd.edu.sg}
\affiliation{
Singapore University of Technology and Design, 8 Somapah Road, Singapore 487372, Singapore
}


\date{\today}

\begin{abstract}
Inertial microfluidics has been limited to dilute particle concentrations due to defocusing (spreading out) at high particle concentrations. We observe a counterintuitive shift of focusing to the outer curved wall under high concentration flow, which contradicts the existing particle focusing theory. We developed a multiphase model incorporating lift forces and particle-particle interactions to explain this behaviour. Numerical simulations validated by experimental data reveal the shift is governed by the ratio of the lift force strength to that of particle interaction frequencies.

\end{abstract}


\maketitle


\section{Introduction} \label{sec:introduction}

Inertial microfluidics is a way of manipulating particles and cells within microchannels in a controlled, deterministic manner, using fluid flow within an intermediate Reynolds number range, typically $1 <$ Re $< 100$. In this flow regime, unlike conventional microfluidics, inertial forces interplay with viscous forces, making lateral migration of suspended particles across streamlines possible. Among existing microfluidic systems, inertial microfluidics has emerged as a promising way for particle and cell separation due to its capabilities in high-volume and high-throughput sample processing \cite{2009_DiCarlo_LabChip, Kuntaegowdanahalli2009, Bhagat2010,2019_Cruz_Hjort_LabChip}.

Most experimental and numerical studies on inertial microfluidics involve particles at dilute concentrations ($<0.5\%$ volume fraction), where they are considered non-interacting, and complex particle-particle interactions can be neglected. Previous work has focused on the interaction between a single particle and the surrounding Newtonian fluid \cite{2009_DiCarlo_Toner_PRL} or only between two particles \cite{2010_Lee_DiCarlo_PNAS}. However, applications in biopharmaceutical manufacturing typically involve high particle or cell concentrations ($>1\%$ volume fraction) to maximize production efficiency \cite{Warkiani2015}. Inertial microfluidics, due to concentration-related non-Newtonian dynamics in such densely-packed regimes (such as in whole blood \cite{2012_Lim_Toner_LabChip}), cannot be adequately explained by dilute-flow physics. Scalable bioprocessing and undiluted whole blood point-of-care applications require accurate quantitative models of inertial microfluidics at higher particle concentrations.

\begin{figure}
    \centering
    \includegraphics{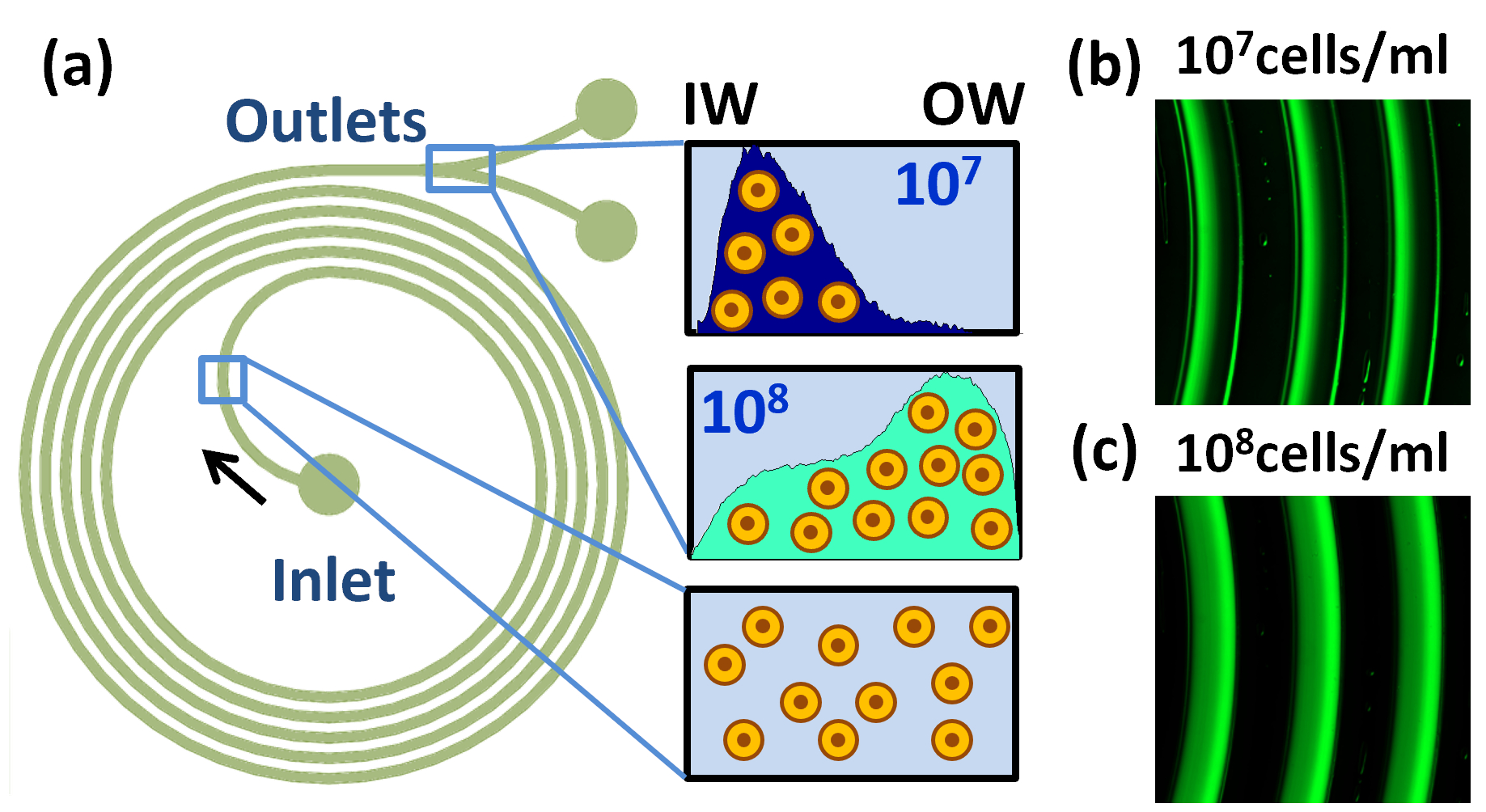}
    \caption{(a) Schematic showing rectangular spiral channel with opposite cell suspension accumulation behavior (b) at the inner wall (IW) when the CHO-GFP cell density is $N = 10^7$ cells/ml and (c) at the outer wall (OW)  when the the CHO-GFP cell density is $N = 10^8$ cells/ml.}
    \label{fig:IW_OW_Focusing}
\end{figure}

For a typical rectangular spiral microchannel (Fig. \ref{fig:IW_OW_Focusing}(a)), particles are known to accumulate at the inner convex wall of a curved channel at low concentrations \cite{Bhagat2009}, as displayed in Fig. \ref{fig:IW_OW_Focusing}(b) for $10^7$ cells per ml of Green Fluorescent Protein-expressing Chinese Hamster Ovary (CHO-GFP) cells, allowing the particles and suspension liquid to be sorted asymmetrically using split channels. This phenomenon is known as inertial focusing. From our previous work \cite{Goh2019}, we demonstrated experimentally that at high concentrations, this behavior inverts, and particles accumulate at the outer concave wall instead, as displayed using $10^8$ cells/ml of CHO-GFP (Fig. \ref{fig:IW_OW_Focusing}(c)). This switching of focusing behaviour at higher cell densities have also been observed by Kwon \textit{et al} \cite{Kwon2021}. They attributed the switching behaviour to the increase in inherent viscoelasticity of the fluid due to the high concentration of cells and compared it to experiments with dilute concentration of particles in solutions with different polymeric additives. 

Here, we explain the switching behaviour using our dense suspension inertial microfluidic particle theory (DENSE-IMPACT) which utilizes a two-phase fluid mixture model for the channel cross-section. In the regime of practical inertial microfluidics, the particle distribution behavior is shown to be governed by the ratio of the lift force to particle interaction frequency contributions. Our numerical results exhibit good correspondence to experimental concentration profiles across a range of flow conditions and channel geometries. Unlike other many-body problems, the low computational cost of this fluid dynamical model without the need for supercomputers affords rapid design and evaluation of microfluidic devices in high concentration regimes, potentially enabling economical and scalable bio-manufacturing processes and medical technology applications.

\section{Theory} \label{sec:theory}
\subsection{Inertial focusing in channels with dilute suspensions}

\begin{figure}
\includegraphics[width = 3.2 in]{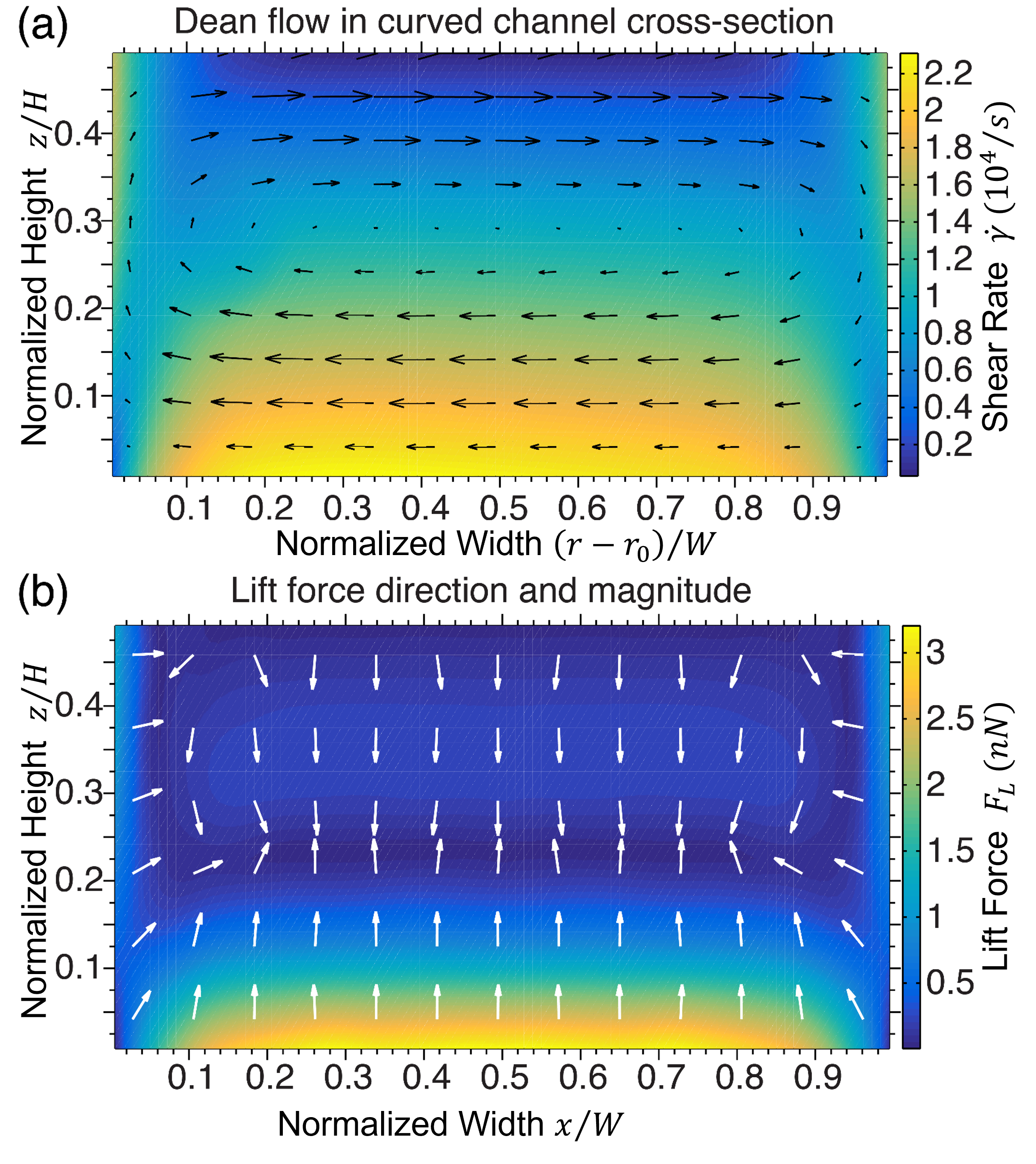} 
\caption{\label{fig:dean_and_lift} The following plots are in the lower half of a rectangular spiral channel with aspect ratio $AR = 5$ and Dean's number $De = 10.7$. $W$ and $H$ are the channel width and height, respectively. $r_0$ is the inner wall radius of curvature. The normalized width is 0 at the inner wall (IW) and 1 at the outer wall (OW) of the rectangular spiral channel. (a) Arrow plot of transverse mixture velocities in the in the limit of zero particle concentration, superimposed on the local shear rate surface plot. The local shear is highest near the walls and lowest at the channel centre. (b) Arrow plot of the lift force direction superimposed on a surface plot of the lift force strength.}
\end{figure}

Segr\'e and Silberberg reported the cross-streamline migration of dilute (volume fraction $\phi<0.4\%$) flowing particles within cylindrical Poiseuille flow (incompressible, steady-state laminar flow) in 1961 \cite{1961_Segre_Silberberg_Nature}. They observed that particles concentrated in a thin annular ring positioned at around 0.6 times the channel radius. Ho and Leal examined the case of two-dimensional Poiseuille flow and quantitatively accounted for the equilibrium annulus and particle trajectories in terms of a balance between two shear-dependent terms: one pointing in the direction of decreasing shear rate (towards the center-line) and another in the direction of increasing shear rate (towards the walls) \cite{1974_Ho_Leal_JFM}. The former term is due to the coupling of the local shear rate with wall corrections, and the latter term is independent of these wall corrections. Schonberg and Hinch derived the first physical theory of inertial migration at finite Reynolds number \cite{1989_Schonberg_Hinch_JFM}, which was later extended to large Reynolds numbers by Asmolov \cite{1999_Asmolov_JFM}.

Di Carlo \textit{et al} explored this phenomenon in straight and curved channels with rectangular cross-sections, observing that with particle volume fractions under $1\%$, particles in straight channels collected in four spots near each of the four wall faces, while particles in curved channels collected in fewer spots \cite{2007_DiCarlo_Toner_PNAS}. They explained the experimental observation in terms of a balance between Dean's drag forces and lift force.

The Dean's drag force arises due to the transverse Dean's flow \cite{1927_Dean_PhilMag,1928_Dean_PhilMag} that occurs with incompressible fluid flow in curved channels. Fig. \ref{fig:dean_and_lift}(a) plots the Dean's flow profile in the lower half of a curved rectangular channel cross-section in the limit of zero particle concentration, where the left boundary corresponds to the inner wall and the right boundary corresponds to the outer wall. The profiles in the upper half channel are the lower half profiles reflected about the horizontal line of symmetry at the normalized height of $0.5$. The Dean's flow takes on a characteristic two-cell configuration, where the flow velocities are the highest at the center and point towards the outer wall, then return along the upper and lower walls of the channel. We also plot the local shear rate $\dot{\gamma}$ of the mixture as the surface plot in Fig. \ref{fig:dean_and_lift}(a). The shear rate has a minimum in the center of the channel and increases towards the wall faces. Since the particle interaction frequency contribution moves particles in the direction of low interaction frequency $\dot{\gamma}\phi$, the shear rate gradient tends to move suspended particles away from the wall faces.

The lift force arises due to the asymmetric flow of fluid around a solid object. The lift force field for a spherical particle in a straight rectangular channel is plotted in Fig. \ref{fig:dean_and_lift}(b). The arrow plot indicates the direction of the lift force, while the surface plot indicates its magnitude. The lift force is strongest along the wall faces with stable equilibrium points directly in front of each wall face. In the limit of low concentration and in straight channels, this lift force contribution dominates and forces suspended particles to occupy positions just adjacent to each wall face, producing four particle accumulation positions centered on each wall face. This four-position focusing geometry was first observed by Di Carlo \textit{et al} in straight square channels and correctly attributed to the lift force contribution \cite{2009_DiCarlo_Toner_PRL}. The lift force completely overwhelms all other forces at the walls, keeping particles away from the walls and ensuring that the particle volume fraction remains near zero at the boundaries.

At low concentrations, the equilibrium position of particles in the channel is governed by the balance of lift forces and Dean's forces. The vector sum of the lift forces and Dean's forces is plotted as the arrow plot in Fig. \ref{fig:OW_explanation}(a). The Dean's forces arise from the Stokes drag associated with the Dean's flow. Particle accumulation occurs at two stable equilibrium positions produced by the superposition of these two contributions, one for each of the upper and lower channel halves. This is the physical basis for inner wall focusing observed by Guan \textit{et al} \cite{Guan2013}.

\subsection{Modeling dense suspensions}

In dense suspensions, when the particle volume fraction becomes large, the rising significance of particle-particle interactions renders the dilute fluid analysis incomplete. Phillips \textit{et al} proposed a model of such interactions using two effects: a spatially varying interaction frequency and spatially varying viscosity \cite{1992_Phillips_Abbott_PhysFluidsA}. Particles that experience more collisions on one side than the other will be displaced away from the region of higher interaction frequency, causing particles to migrate away from regions of high interaction frequency. Similarly, when flowing in a region of spatially varying viscosity, particles may be deflected towards the region of lower viscosity. However, to our knowledge, these particle-particle interactions have not been examined in the context of inertial focusing in a microchannel \cite{Bazaz2020}.

\subsection{Multiphase fluid dynamical model}

In this study, we consider a steady-state, two-dimensional, two-phase model for the rectangular spiral channel cross-section, treating the suspended particles as part of a continuous dispersed phase field and introducing a diffusion equation for concentrated suspensions, as proposed by Phillips \textit{et al} \cite{1992_Phillips_Abbott_PhysFluidsA}. The dispersed phase field is parameterized by a spatially varying volume fraction, which represents the local volume fraction occupied by the suspended particle phase. In moving from finite-sized particles to the continuous phase approximation, we lose information on the lift forces acting on the dispersed phase, since lift forces arise due to flow perturbation around finite-sized objects. We compensate for the lack of lift forces by introducing an external position-dependent lift force field acting on the dispersed phase. The lift force field is computed in a separate fluid dynamical simulation in which we place a freely rotating sphere at different positions in a straight channel with the same cross-section and explicitly compute the lift force generated by laminar flow across the sphere.

Consider neutrally-buoyant particles of radius $a$ in a fluid of density $\rho$ and dynamic viscosity $\eta_f$, with a mixture velocity field $\vec{u}$ and pressure field $p(\vec{r})$. Let the dimensionless dispersed phase volume fraction be $\phi$. Since the dispersed phase is neutrally buoyant, this volume fraction is identical to the dispersed phase mass fraction. The average volume fraction of the mixture is $\phi_{avg}$, which is proportional to the average number density or concentration $N$ of particles: $\phi_{avg}=(4/3)\pi a^3 N$.

The numerical model is implemented using the Multiphase Mixture, Laminar Flow and General Form PDE modules in COMSOL 5.2a. The steady-state incompressible mixture momentum balance equation, under the laminar flow condition, is \cite{1996_Manninen_Kallio}:
\begin{eqnarray}
    \rho (\vec{u}\cdot\nabla)\vec{u} =\vec{f} -\nabla p + \nabla\cdot[\eta(\nabla\vec{u}+(\nabla\vec{u})^T)] \nonumber \\
    - \nabla\cdot [\rho\phi(1-\phi)\vec{u}_{slip}\vec{u}_{slip}^T] \label{eqn:transport}
\end{eqnarray}

$\vec{f}$ is the volumetric forces acting on the mixture, $p$ is the pressure, and $\rho$ is the constant density. The slip velocity $\vec{u}_{slip}=\vec{u}_d - \vec{u}_f$ is the difference between the dispersed phase velocity $\vec{u}_d$ and the fluid phase velocity $\vec{u}_f$. 

The mixture viscosity $\eta$ depends on the local dispersed phase volume fraction $\phi$ through the Krieger-Dougherty relation \cite{1959_Krieger_Dougherty_TransSocRheology}:
\begin{equation}
    \eta(\phi) = \eta_f \left(1-\frac{\phi}{\phi_{max}}\right)^{-2.5\phi_{max}}
\end{equation}

$\phi_{max}$ is the maximum allowed dispersed phase volume fraction. We use the fitted value of $\phi_{max}=0.68$ based on experiments at the high-shear limit for a suspension of hard spheres \cite{1972_Krieger_AdvColloid}. The viscosity thus increases from the nominal fluid phase viscosity $\eta_f$ at zero volume fraction to infinity as $\phi$ approaches $\phi_{max}$.

The mixture velocity $\vec{u}$ is the volume-fraction-weighted velocity of the dispersed phase and fluid phase:
\begin{equation}
    \vec{u} = \phi\vec{u}_d + (1-\phi)\vec{u}_f
\end{equation}

The slip velocity $\vec{u}_{slip}=\vec{u}_d-\vec{u}_f$ is key to introducing particle-particle interactions. We write it with three contributions:
\begin{equation}
(1-\phi)\vec{u}_{slip} = -K_c a^2\nabla(\dot{\gamma}\phi) - K_\eta a^2\phi\dot{\gamma}\nabla\eta/\eta+f_h \vec{u}_{st} \label{eqn:slip}
\end{equation}

$\dot{\gamma}$ is the magnitude of the shear rate tensor, which captures the local velocity gradient. The term proportional to $K_c$ represents the effect of a spatially-varying interaction frequency, and the term proportional to $K_\eta$ represents the effect of a spatially-varying viscosity. The interaction frequency is proportional to the particle relative motion and the concentration of other particles, thus is proportional to the product of shear rate and local particle volume fraction: $\dot{\gamma}\phi$. $K_c$ and $K_\eta$ are dimensionless empirical fitting parameters and are the only two degrees of freedom in our model.

The third term in Equation \ref{eqn:slip} represents the effect of settling velocity $\vec{u}_{st}$ due to interphase forces that occur independently of gradients in interaction frequency and viscosity. It was introduced by Zhang and Acrivos \cite{1994_Zhang_Acrivos_IntJMultiphaseFlow}. $f_h$ is a hindering function that accounts for the decrease in settling velocity at high concentrations:
\begin{equation}
    f_h = \frac{\eta_f(1-\phi_{avg})}{\eta}
\end{equation}

We introduce lift forces through the secondary flow term in the dispersed phase flux equation, relating the secondary flow velocity $\vec{u}_{st}$ to the force term by assuming that the lift forces induce Stokes flow in the absence of concentration gradients.
\begin{equation}
    \vec{u}_{st} = \frac{\vec{F}_L}{6\pi \eta_f a}
\end{equation}

This is the key step that enables the concentrated suspension model to be adapted for finite-sized particles. Fig. \ref{fig:dean_and_lift}(b) exhibits the direction and magnitude of the lift force across the channel. We neglect other volumetric forces. The model is solved iteratively for the steady-state mixture velocity profile and volume fraction $\phi(\vec{r})$.

\section{Results}\label{sec:results}
\subsection{Experimental measurement of concentration profiles and model fitting}

\begin{figure*}

\includegraphics[width = 6 in]{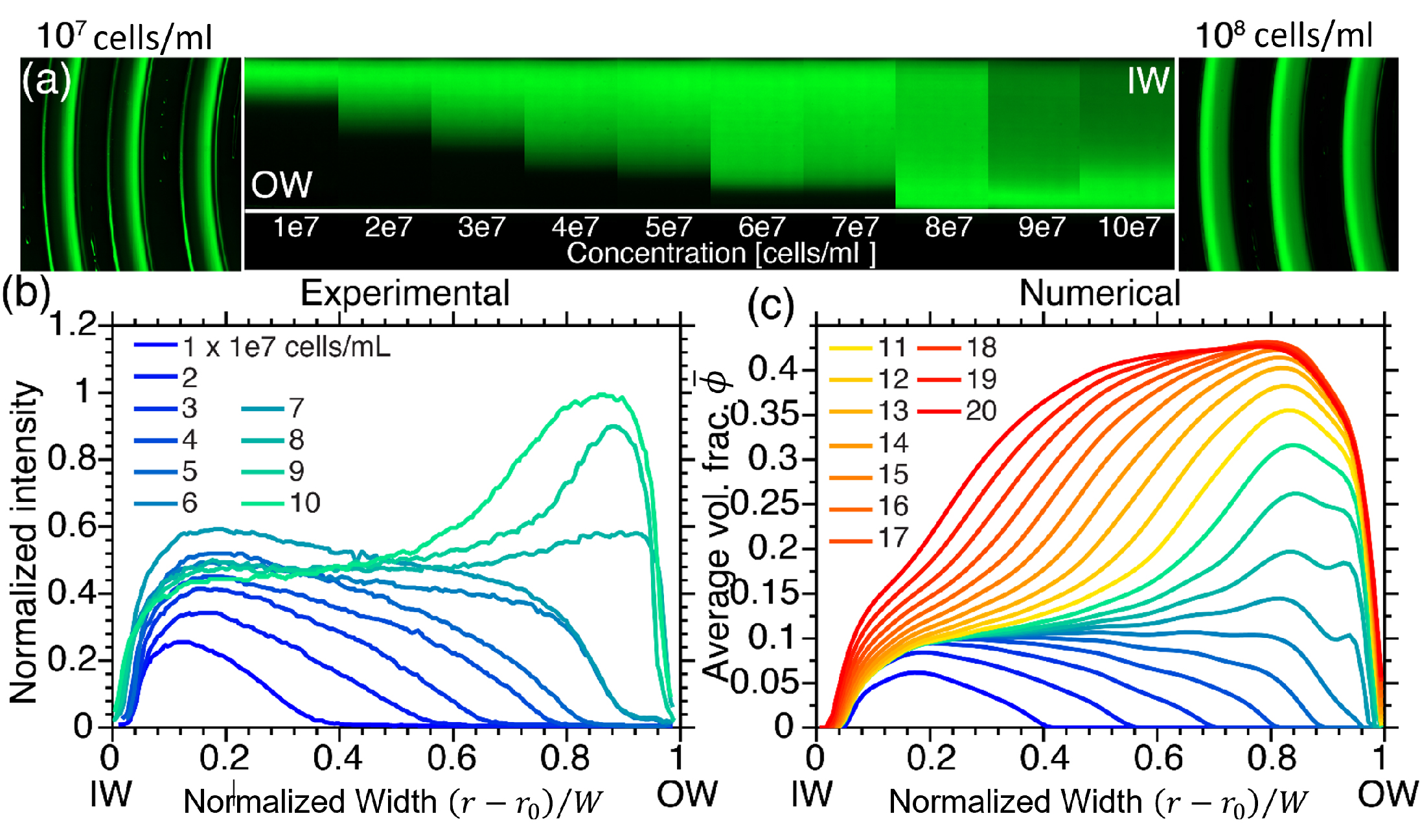}
\caption{\label{fig:concsweep} (a) Microscope image of fluorescent CHO-GFP cells accumulating at the inner wall (IW) of the spiral rectangular channel when the concentration is $10^7$ cells/ml (extreme left) to the outer wall (OW) of the spiral rectangular channel when the concentration is $10^8$ cells/ml (extreme right). Each image is normalized to have the same maximum pixel intensity value. (b) Experimental plot showing the intensity profile peak position migrating from IW to OW as the cell concentration is increased from $10^7$ to $10^8$ cells/ml. (c) The numerically-modelled particle volume fraction profile for the experimental concentrations and beyond shows very close correspondence. }
\end{figure*}

Inertial focusing at varying cell densities is studied by flowing CHO-GFP cells, with radius $a=7-8\mu m$, into rectangular spiral microchannels with different aspect ratios and radii of curvature. Fluorescent cell streaks are observed from the top-down with an optical microscope. Fig. \ref{fig:concsweep}(a) exhibits the cross-channel fluorescent intensity distribution captured as the CHO-GFP cell number density is increased from $10^7$ cells/ml ($1.76\%$ volume fraction) to $10^8$ cells/ml ($17.6\%$ volume fraction) in a rectangular spiral microchannel with channel aspect ratio $AR = W/H = 5$, where W = width and H = height of the rectangular cross-section, and Dean's number $De=Re_C \sqrt{D_h/2R}=10.7$). The intensity profiles are normalized so that the area under each curve is proportional to the input concentration. At $10^7$ cells/ml, the cells accumulate near the inner (concave) wall of the channel, but as the concentration increases to $10^8$ cells/ml, the focus shifts to the outer (convex) wall. The cross-sectional intensity profiles are overlaid in Fig. \ref{fig:concsweep}(b), demonstrating a persistent and relatively uniform cell distribution in the inner half channel at concentrations exceeding $6\times 10^7$ cells/ml.

The experimental concentration profiles in Fig. \ref{fig:concsweep}(b) were used to select the numerical values of $K_c$ and $K_\eta$ in our model. Since the channels were viewed from above, the fluorescence intensity measurements correspond to the average cell concentration of each vertical section, less any line-of-sight obstruction. The best values of the empirical parameters based on parameter sweeps are $K_c=0.18$ and $K_\eta=0.20$. For comparison, the values fitted by Phillips \textit{et al} \cite{1992_Phillips_Abbott_PhysFluidsA} are $K_c=0.41$ and $K_\eta=0.62$. The difference between the two estimates is largely due to the experiments being in different concentration regimes; Phillips examines systems with volume fractions in excess of 50\% with a model that is less accurate for local volume fractions less than 40\%, while our experiments span a lower concentration regime, below 20\%. These systems also have different geometries: Phillips examines Couette flow between parallel surfaces, whereas we study a finite rectangular spiral channel. More complex concentrated suspension models have allowed the empirical parameters to vary based on the average \cite{1998_Tetlow_Altobelli_JRheology} and local \cite{2002_Rao_Altobelli_IntJNumMethodsFluids} volume fractions, but we find that simple constant values for $K_c$ and $K_\eta$ are sufficient to reproduce our experimental observations. 

Fig. \ref{fig:concsweep}(c) exhibits the numerical cross-channel concentration profiles (in blue-green curves) for an effectively identical channel geometry ($AR = 5$) as in Fig. \ref{fig:concsweep}(b). The numerical profiles exhibit good correspondence to the experimental profiles but predicts a secondary intensity peak near the outer wall of the channel, which is not observed in the experiment. This discrepancy may be due to channel imperfections, such as a slightly-sloping channel, which causes the channel to deviate from a perfect rectangular cross-section. The outer wall peak is observed to continue growing in height beyond $10^8$ cells/ml as the concentration increases until the concentration reaches around $1.5\times 10^8$ cells/ml, at which the maximum local volume fraction achieved in the channel reaches 60\%, which is close to the model maximum of 68\% (as $\phi_{max}=0.68$). Beyond this point, the cell distribution broadens back in the direction of the inner wall, eventually dispersing uniformly over the channel cross-section. 

The choice of the $K_c$ and $K_\eta$ fitted parameters was validated by comparing the model predictions to experimental concentration profiles under different suspension flow rates and channel aspect ratios. The model concentration profiles agree well with the experimental measurements, as detailed in Section S4 of the Supplementary Information. 

\section{Discussion} \label{sec:discussion}
\subsection{Mechanism for concentration-driven outer-wall focusing}
\begin{figure*}
\includegraphics[width = 6 in]{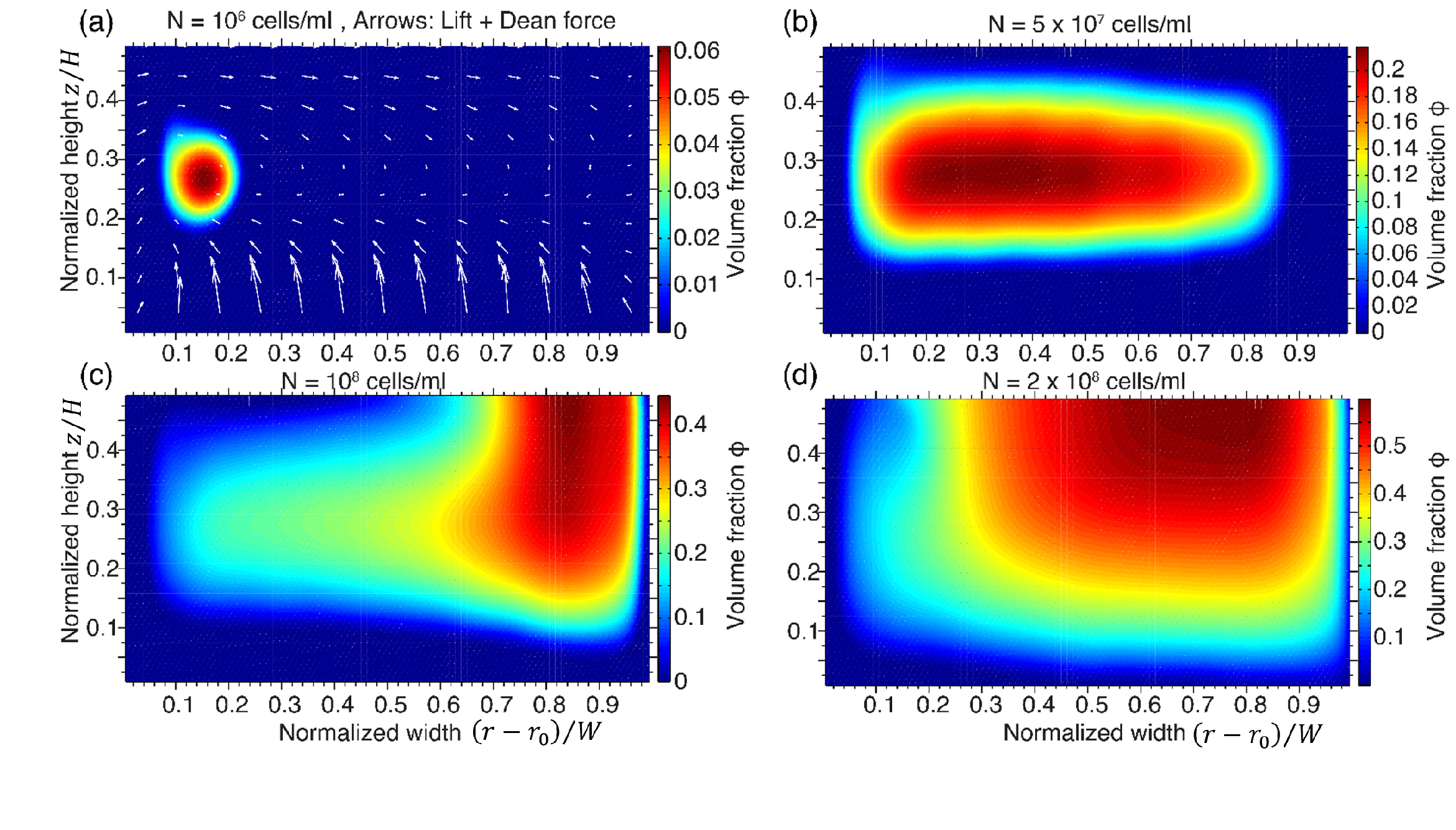} 
\caption{\label{fig:OW_explanation} Simulated particle volume fractions in the lower half of a rectangular spiral channel, with aspect ratio $AR = 5$ and Dean number $De = 10.7$. The normalized width is 0 at the inner wall (IW) and 1 at the outer wall (OW) of the rectangular spiral channel. (a) Arrow plot: combined lift and Dean force. Surface plot: particle volume fraction, for a low input concentration of $10^6$ cells/ml. The particles congregate at the stable minima of the combined lift and Dean force. (b) As the input concentration increases to $5\times 10^7$ cells/ml, the inter-particle interaction frequency gradient pushes the distribution towards the central region of low shear rate. (c) As the input concentration increases to $10^8$ cells/ml, the particles pile up at the outer wall. (d) At even higher input concentrations of $2\times 10^8$ cells/ml, the interparticle forces are able to fill up the inner channel as well.}
\end{figure*}

The two-dimensional transverse concentration profile obtained through numerical predictions affords us a mechanism for the experimentally-observed concentration-driven outer wall focusing. Fig. \ref{fig:OW_explanation} plots the cross-sectional particle volume fraction profiles at four concentrations. At low concentrations ($10^6$ cells/ml), particles accumulate at the minima of the lift and Dean's force (Fig. \ref{fig:OW_explanation}(a)). As the particle concentration increases to $5\times 10^7$ cells/ml (Fig. \ref{fig:OW_explanation}(b)), the particle distribution broadens in the direction of reduced shear rate (i.e., to the channel centre) as particles migrate to regions of lower collision frequency, inducing concentration-driven redistribution to the outer wall. The particle distribution still remains in the region of low lift forces and Dean’s flow velocities.

As the particle concentration increases further to $10^8$ cells/ml (Fig. \ref{fig:OW_explanation}(c)), the particle distribution spreads in the vicinity of the outer wall and cannot progress further due to the increase in shear rate and opposing lift force there. Particles are prevented from migrating in the direction of the bottom of the channel by the upward-pointing lift force acting on particles there, which overwhelms other contributions. Particles thus accumulate along the horizontal axis of symmetry, and thus lie in the region of rapid outward-flowing Dean's flow. The rapid Dean's flow biases the particle accumulation towards the direction of the outer wall, producing an outer wall particle accumulation peak. 

At even higher particle concentrations, the local particle volume fraction approaches the maximum value allowed ($\phi_{max}=0.68$). Since the mixture viscosity increases very rapidly near $\phi_{max}$, the spatially-varying viscosity contribution dominates in this regime, pushing the particle distribution outwards towards the regions not yet occupied by particles. The close-packing of the particles increases the contribution of the $K_c$ and $K_\eta$ terms, which partially overcomes the outward-pointing Dean's flow along the horizontal axis of symmetry and the upward-pointing lift force near the base/top of the channel. The outer wall peak thus broadens back in the direction of the inner wall, while extending slightly into the vertical extremes of the channel. Fig. \ref{fig:OW_explanation}(d) exhibits the broad particle distribution when the initial concentration is $2\times 10^8$ cells/ml (35.2\% volume fraction).

\subsection{Scaling law for inner and outer wall accumulation}
We parametrize the particle distribution with the inner half flux fraction, which is the percentage of particle flux contained within the inner half of the channel (i.e., with normalized width $<0.5$). This fraction is 0 under pure outer wall accumulation and 1 in pure inner wall accumulation. 

\begin{figure}
\includegraphics[width = 3 in]{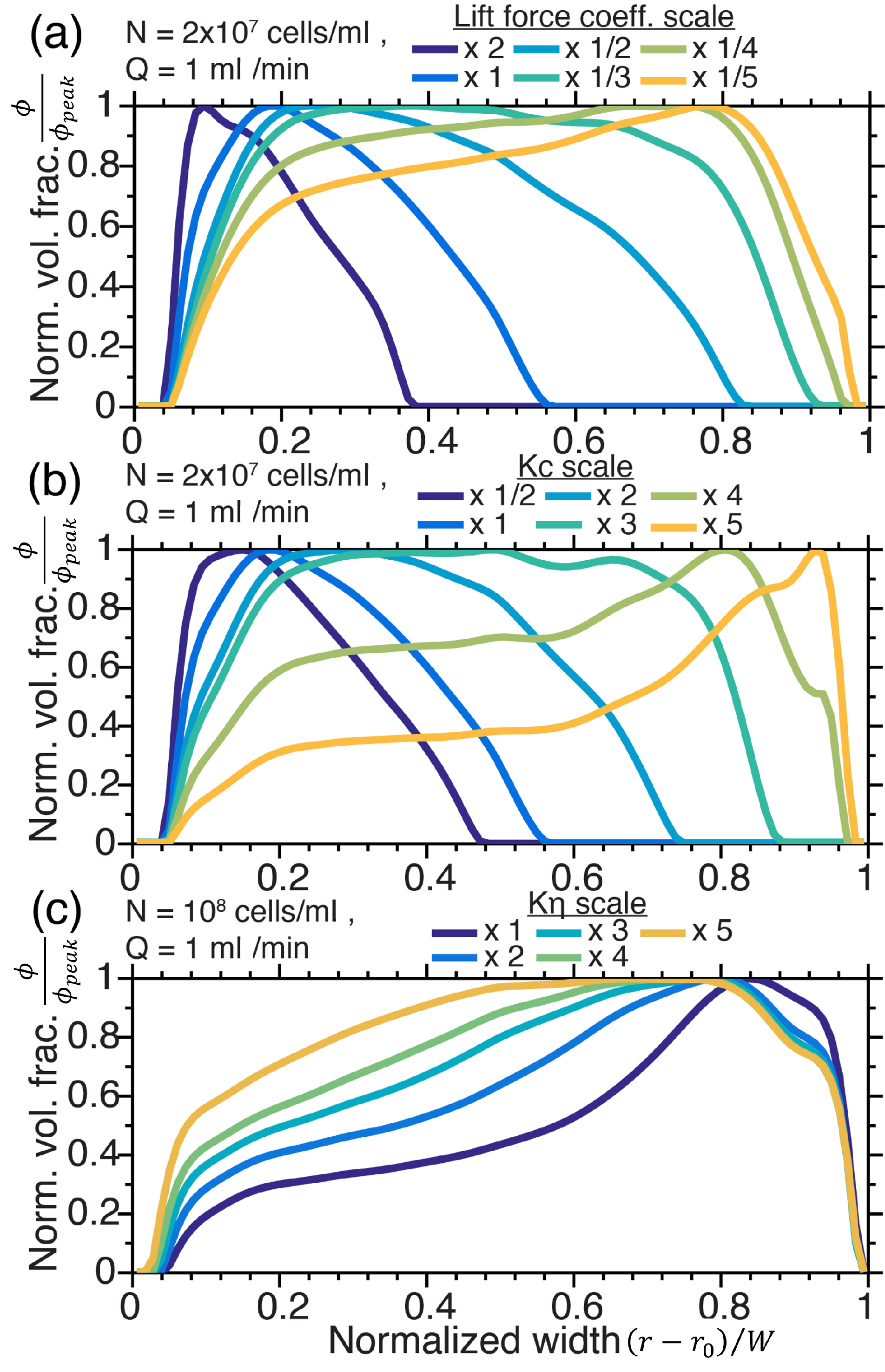} 
\caption{\label{fig:D1D2FL} (a) Effect of varying the lift force coefficient on the normalized volume fraction. (b) Effect of varying the magnitude of $K_c$, which controls the spatially-varying interaction frequency contribution. As the lift force is decreased or $K_c$ is increased, the particle distribution is observed to migrate from IW to OW accumulation. (c) Effect of varying the magnitude of $K_\eta$, which controls the spatially-varying viscosity contribution. The viscosity contribution only plays a significant role at high volume fractions ($>10\%$ volume fraction) due to the nonlinear volume fraction-dependence of the viscosity.}
\end{figure}

To identify the key driving factors behind the inner-outer wall transition, we sweep the magnitude of the various physical contributions and examine their effects on the model profile predictions in Fig. \ref{fig:D1D2FL}. We vary the magnitudes of the lift force, $K_c$, $K_\eta$, and plot the predicted channel concentration profiles in Fig. \ref{fig:D1D2FL}(a), (b), and (c), respectively. We observe that the transition from inner wall to outer wall focusing can be induced by decreasing the value of the lift force coefficient or increasing $K_c$ while keeping all other parameters constant (Fig. \ref{fig:D1D2FL}(a-b)). Increasing $K_\eta$ has the only effect of broadening the focusing peak (Fig. \ref{fig:D1D2FL}(c)), as expected since it induces particle flow down viscosity gradients from crowded regions to less crowded regions. However, changes in $K_\eta$ only have significant effects at high volume fractions in excess of 10\%. This is due to the nonlinearity of the viscosity-volume fraction relationship: the viscosity changes slowly as a function of particle volume fraction at low volume fractions, but this dependence increases exponentially near $\phi_{max}$. Since the inner-outer wall transition occurs at low volume fractions less than 10\% as well, the $K_\eta$ effect cannot be the main driving factor behind the transition.

The similar concentration profile shifts obtained under the decrease in lift force strength and increase in $K_c$ influence suggests that the accumulation dynamics depends on the ratio between the lift force strength and spatially-varying interaction frequency strength. We derive the scaling behavior of the four primary force contributions in Supplementary Information section S2. The ratio $K$ between the lift force $F_L$ and $K_c$ force term $F_c$ scaling is:
\begin{equation}
K = \frac{F_L}{F_c} = \frac{U}{a^2N} = \frac{Q}{a^2WHN}
\end{equation}

\begin{figure}
\includegraphics[width = 3.3 in]{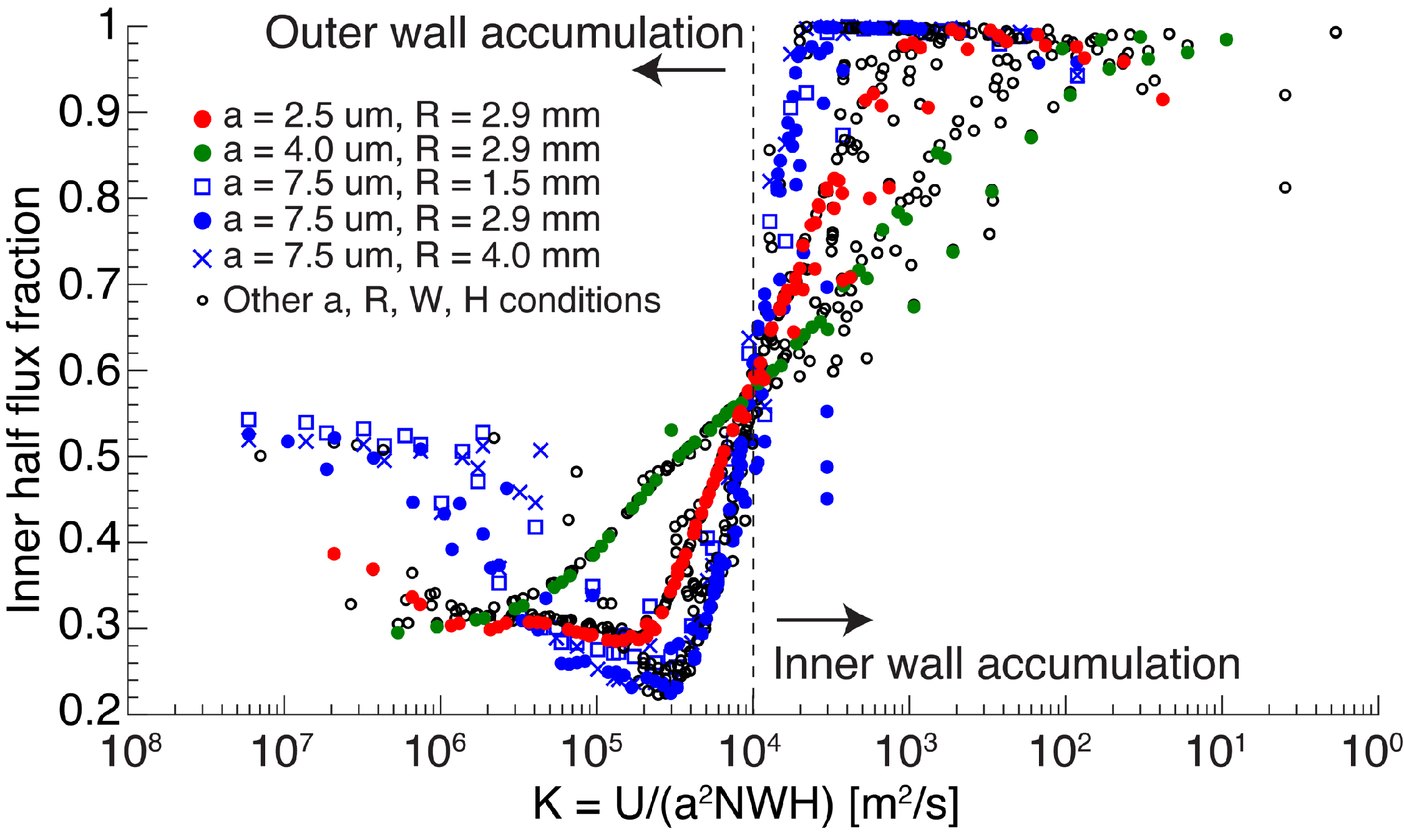} 
\caption{\label{fig:collapsed_plot} Numerical predictions for the dependence of the inner half flux fraction (fraction of particle flux located in the inner half of the channel) on the parameter $K=Q/(a^2 NWH)=U/(a^2 N)$. The data points for different particle and channel regimes collapse along master curves. $K=10^{-4}$ provides a convenient cut-off separating the outer wall accumulation and inner wall accumulation regimes. Systems with other parameter values also obey the critical $K$ cutoff; these are plotted as empty black circles and span: $10^{-3}$ ml/min⁡ $<$Q$<$2 ml/min, 0.5 mm $<$R$<$ 46 mm,$50 \mu m<H<140 \mu m,300 \mu m < W < 500 \mu m$ and $2 \mu m<a<10 \mu m$.}
\end{figure}

$Q$ is the volumetric flow rate and $U$ is the mean flow velocity. We observe that systems with the same value of $Q/N$ and with all other parameters constant share virtually the same predicted particle distribution (Supplementary Information Section S5). Furthermore, when the inner half flux fractions for three particle sizes ($a=2.5 \mu m$ (for \textit{Pichia pastoris} yeast), $4.0 \mu m$ (for red blood cells), and $7.5 \mu m$ (for CHO cells) are plotted against $K$, the data points for various $(Q,N)$ combinations collapse along master curves that intersect at an inner half flux fraction of 0.55 (Fig. \ref{fig:collapsed_plot}). Changing the value of $R$ does not affect the accumulation behaviour substantially. The blue squares and crosses in Fig. \ref{fig:collapsed_plot} represent the inner half flux fractions of systems with innermost radii values of 1.5 mm and 4.0 mm, respectively. Systems with different $R$ values still fall along the same blue master curve for $a=7.5 \mu m$. The inner half flux fraction intersection point at $K=10^{-4}m^2/s$ of 0.55 serves as a natural boundary for us to classify inner and outer wall accumulation. Systems with $K<10^{-4} m^2/s$ are predicted to have inner half flux fractions of less than 0.55, which indicates outer wall accumulation, and vice versa. In the experimental system exhibited in Fig. \ref{fig:concsweep}, the $K$ value decreases from $6\times 10^{-4} m^2/s$ to $6\times 10^{-5} m^2/s$, as the concentration increases from $10^7$ cells/ml to $10^8$ cells/ml, crossing the $K$ threshold and thereby shifting from an inner wall focusing condition to the outer wall focusing condition.

This critical value for $K$ in determining inner and outer wall focusing holds for a wide range of channel geometries and particle sizes. The empty black circles in Fig. \ref{fig:collapsed_plot} represent other systems with different flow rates, radii of curvature, aspect ratios, and particle diameters that are typical for inertial microfluidics. The lack of dependence of $K$ on the channel radius of curvature $R$ indicates that changes in the Dean’s flow contribution are not significant in determining the particle focusing distribution within the channel. However, the Dean's contribution is still important in accounting for the direction of particle build-up by pushing particles near the horizontal axis of symmetry towards the outer wall. The Dean's contribution breaks the left/right symmetry of the system and thus allows for particle distributions that are skewed to the inner or outer walls. 

\section{Conclusion}

Based on an experimental and numerical study of the inertial dynamics of concentrated suspensions in curved channels, we have made the first observation of outer-wall particle accumulation under high particle volume fractions. The particle accumulation effect may be exploited in a split channel junction to remove less-concentrated liquid from concentrated suspensions, which is useful for extracting dissolved biopharmaceutical products generated by suspended cells in a bioreactor. The strategy of incorporating particle-particle interactions and lift forces in dispersed phases may bring additional insight in modeling other dense suspension systems beyond bio-manufacturing, from soil rheology to wet granular flows.


\section{Acknowledgments}
 The authors thank Nikon Imaging Centre Singapore for confocal microscope use. This work is supported by GAP funding from A*STAR Singapore. S.W.D.L. is supported by the Schmidt Science Fellows, in partnership with the Rhodes Trust. S.W.D.L. is also supported by A*STAR Singapore through the National Science Scholarship Scheme.

\bibliography{references}

\newpage

\renewcommand{\theequation}{S\arabic{equation}}
\renewcommand{\thesection}{S\arabic{section}}  
\renewcommand{\thefigure}{S\arabic{figure}}

\textbf{Supplementary Information}

\section{Numerical methods}

For all simulations, we use the water density $\rho =1000 kg m^{-3}$ for both the fluid and dispersed phase, and the water dynamic viscosity $\eta_f=8.9\times 10^{-4}$ Pa s for the fluid phase. 

\subsection{Lift force calculation}

We calculate the lift force field $\vec{F}_L$ separately, then export the values so that the multiphase simulations can draw upon the same lift force information. The lift force calculation strategy was adapted from Di Carlo \textit{et al} \cite{2009_DiCarlo_Toner_PRL}.

We place a spherical void of radius $a$, representing a spherical particle, in a rectangular domain of x-width $W=500 \mu m$, y-height $H=100 \mu m$ and z-length $2000 \mu m$. The void is centred at the $z=0 \mu m$ plane of the length dimension, and its position is varied in one $xy$ quadrant of the centre plane. 

We are interested in the transverse forces acting on the particle when it translates in a drag-free and zero-torque condition. To implement this condition, we introduce four degrees of freedom: $v_p,\omega _x,\omega_y,\omega_z$, which represent the axial translational velocity of the particle and the three components of its angular velocity vector, respectively. We then introduce four global equations that the simulation solution has to satisfy in this drag-free zero-torque condition:
\begin{eqnarray}
    F_z =  0\\
    \tau_x = 0 \\
    \tau_y = 0 \\
    \tau_z = 0
\end{eqnarray}

where $F_z$ is the axial force on the particle (drag), and the $\tau_i$ parameters are the net torques on the particle with respect to its geometrical center. 

To compute the drag force in the z-direction, $F_z$, we integrate the z-component of the total stress over the sphere surface. To compute the three torque values, we calculate the differential forces $d\vec{F} = \overline{K}\hat{n}$ over the sphere surface, where $\overline{K}$ is the viscous stress tensor and $\hat{n}$ is the normal vector to each mesh element. We then integrate the differential torques over the sphere surface to obtain the total torque:
\begin{equation}
    \vec{\tau} = \iint_S d\vec{\tau} = \iint_S \vec{r}\times d\vec{F}
\end{equation}

We adopted a particle-centric frame of reference, and set the particle to be stationary in our frame, while the four walls of the channel in the width and height dimensions were set to velocity boundary conditions in the length direction to $-v_p \hat{z}$.
 
At one end of the channel in the length dimension, which we call the inlet, we introduce a fully-developed laminar flow profile (flow rate Q=1 ml/min) respecting the no-slip conditions at the wall boundaries. We explicitly set the transverse velocity components to zero and only introduce the velocity component normal to the inlet surface. We subtract the particle axial translational velocity $v_p$ from this laminar flow profile $\vec{v}_{laminar}$ to give the fluid velocity boundary condition at the inlet. 
\begin{equation}
    \vec{v}_{inlet} = \vec{v}_{laminar} - v_p\hat{z}
\end{equation}

At the other end of the channel, we impose a zero pressure boundary condition without imposing normal flow or suppression of backflow. 

We also allow the particle to rotate freely, and incorporate this motion by imposing a velocity boundary condition at the void boundary:
\begin{equation}
    \vec{v} = \vec{\Omega}\times(\vec{r}-\vec{r}_0), \quad |\vec{r}-\vec{r}_0|=a
\end{equation}

where $\vec{\Omega}=(\omega_x,\omega_y,\omega_z)$ is the angular velocity of the particle and $\vec{r}_0$ is the position of the void center. 

The experimental geometry is meshed using a boundary layer mesh with 5 boundary layers and a boundary layer stretching factor of 1.2. The thickness of the first layer at the void boundary is set to $a/10$ and the thickness of the first layer at the channel walls is set to $1 \mu m$. The maximum element size at the spherical void boundary is set to $0.6 \mu m$, which is much smaller than the sphere radius $a$, in order to finely resolve the flow near the particle. The volume tetrahedral elements are allowed to grow in size away from the particle at a maximum element growth rate of 1.13 to a maximum element size of $20 \mu m$. The minimum element size of any volume or fluid element is set to $0.01 \mu m$. The maximum surface element size on the inlet and outlet is fixed at $5 \mu m$.

We solve for the fluid velocity and pressure using the incompressible laminar flow equations:
\begin{eqnarray}
    \rho(\vec{u}\cdot\nabla)\vec{u} = \nabla\cdot [-p\overline{I}+\eta(\nabla\vec{u}-(\nabla\vec{u})^T)] \\
    \nabla\cdot\vec{u}=0
\end{eqnarray}

We solve the system using a fully coupled direct solver with a relative tolerance of $10^{-6}$. Upon convergence, we calculate the net transverse force acting on the particle by integrating the $x$ and $y$ components of the total stress over the sphere surface.

We use the leading order term from Hood \textit{et al} \cite{2015_Hood_Roper_JFM} to calculate the lift coefficient $\vec{c}_4$ associated with the lift forces. That is, we write the lift forces with the following scaling relationship:
\begin{equation}
    \vec{F}_L = \vec{c}_4 \frac{\rho U^2a^4}{D_h^2}
\end{equation}

where $U=Q/(WH)$ is the mean channel velocity and $D_h=2WH/(W+H)$ is the hydraulic diameter. $\vec{c}_4$ is the dimensionless vectorial lift coefficient which depends on the channel geometry alone.

To obtain the lift coefficients in the other three quadrants of the channel, we exploit the vertical and horizontal planes of symmetry along the channel and reflect the $\vec{c}_4$ vectors about these planes. In order to respect this reflection symmetry, we force the components of $\vec{c}_4$ normal to these symmetry planes to be exactly zero along these planes. 

We use 7 linearly-spaced channel height positions and 10 linearly-spaced channel width positions in this study. These 70 datapoints are used to perform a spline interpolation to express the lift coefficient components as a $C^2$ function (exported as a $200 \times 500$ element matrix) over the channel cross-section. Points lying outside the region sampled are linearly extrapolated. We also explicitly calculate the numerical first and second derivatives for each point in the interpolated dataset and import these derivative values into the COMSOL simulation as distinct interpolated functions.

\subsection{Multiphase computational fluid dynamics}

The multiphase flow simulations were performed using the Multiphase Mixture, Laminar Flow and General Form PDE modules in COMSOL 5.2a controlled using LiveLink for MATLAB R2013b.

We use cylindrical coordinates $(r,\theta,z)$ and consider the plane where $\theta=0$. In this plane, we take the rightward direction to be $\hat{r}$ the direction into the plane away from the viewer to be $\hat{\theta}$ and the upward direction to be $\hat{z}.$ Let the position vector be $\vec{r}$ and let the scalar components of the velocity vector be $\vec{u}=(u(\vec{r}),v(\vec{r}),w(\vec{r}))$.

We solve for the steady-state distribution of fluid and dispersed phase variables for a mixture flowing in a rectangular toroid (channel width $W$ and height $H$), subject to rotational symmetry in the azimuthal direction. This choice of simulation geometry allows us to simulate the field parameters in a rectangular 2D cross-section of the toroid instead of the fully 3D spiral geometry. This simulation is performed using the swirl flow condition in the Mixture model, which introduces the azimuthal components normal to the cross-sectional area. The inner wall is placed at a radial position of $R$ and the outer wall is located at $R+W$. The base of the channel is located at $z=0$ and the horizontal symmetry axis is located at $z = H/2$. 

We use the General Form PDE module to introduce the local shear rate $\dot{\gamma}$ as a field over the 2D surface using quadratic Lagrange elements. In an axisymmetric system where the $\theta$ derivatives vanish, the local shear rate can be explicitly written in terms of the mixture velocity components:
\begin{widetext}
\begin{equation}
    \dot{\gamma} = \sqrt{2\left(\frac{\partial u}{\partial r}\right)^2 + 2\left(\frac{\partial w}{\partial z}\right)^2 + \left(\frac{\partial u}{\partial z}+\frac{\partial w}{\partial r}\right)^2  + \left(\frac{\partial v}{\partial r}\right)^2 + \left(\frac{\partial v}{\partial z}\right)^2}
\end{equation}    
\end{widetext}

The fluid velocity components, pressure, and dispersed phase volume fraction are represented using linear Lagrange elements. This linear approximation improves the stability of the numerical simulation against random fluctuations in volume fraction that result in divergence. 
We use a mapped mesh to divide the 2D rectangular cross-section into identical rectangles. 

We impose no-slip boundary conditions for the mixture velocity along the channel base and sides, and a symmetry condition for the mixture velocity along the horizontal symmetry axis. We impose zero volume fraction conditions for the dispersed phase along the channel base and sides and a symmetry condition along the horizontal symmetry axis. In our regime of interest, the zero volume fraction condition yields the same solution as the zero dispersed phase flux condition, but provides better numerical stability since transient parameter singularities tend to occur along the channel boundaries. Forcing the dispersed phase volume fraction to zero along these boundaries hence smooths over these singularities and improves convergence.

We impose a zero pressure point constraint at the upper corner of the inner wall.

We constrain the total mixture volumetric flux through the 2D surface to be equal to the specified volumetric flux $Q$ using a global constraint on the axial velocity component $v(r,z)$. The factor of 2 arises because we are only considering half the channel cross-section.
\begin{equation}
    \iint_S v dr dz = \frac{Q}{2}
\end{equation}

We also constrain the dispersed phase flux through the surface through the surface using another global constraint on both the dispersed phase volume fraction and the axial velocity:
\begin{equation}
    \iint_S v\phi drdz = \frac{Q\phi_{avg}}{2}
\end{equation}

We solve the system using a fully coupled direct solver with a relative tolerance of $10^{-3}$.

We introduce penalty diffusion for the dispersed phase volume fraction to suppress negative volume fractions. To remove the negative volume fractions that still remain after convergence, which are only on the order of $10^{-4}$, we rescale the volume fraction values between the minimum volume fraction obtained for that run and the $\phi_{max}$ value of 0.68:
\begin{equation}
    \phi_{scaled} = \frac{\phi-\min(\phi)}{\phi_{max}-\min(\phi)}\phi_{max}
\end{equation}

This ensures that the scaled volume fraction lies between 0 and $\phi_{max}$.

In comparing the model predictions with that obtained from experimental profiles, we find that the best values to use for the $K_c$ and $K_\eta$ parameters are $K_c=0.18,K_\eta=0.20$.

We perform a mesh invariance study to ensure that the solutions are properly converged and to find the minimum degrees of freedom necessary to capture the dynamics. We chose an optimal mesh rectangular element size of $2.2 \mu m\times 3.3 \mu m$ (width direction × height direction). For a system where $N = 10^8 \text{ cells/ml}, W=500 \mu m,H=100 \mu m,R=2.9 \text{ mm},Q=1$ ml/min, this mesh resolution choice produces a maximum element-wise volume fraction error of 12\% and an RMS error of 2.4\%, for approximately 63 000 degrees of freedom and a solution time of 280 core-minutes and using less than 6 GB of RAM. 

\section{Scaling relationships} \label{sup_sec:scaling}

In this section, we examine the scaling relationships for each of the effects that act on the dispersed phase. There are four forces to consider: lift forces, Dean’s forces, and the associated contributions of the spatially-varying interaction frequency ($K_c$ parameter) and viscosity ($K_\eta$ parameter). 

The leading order scaling for lift forces in straight, two-dimensional Poiseuille flow (in powers of the ratio of the particle size $a$ to the channel width $l$) has been derived by Ho and Leal \cite{1974_Ho_Leal_JFM} to be:
\begin{equation}
    F_L = \frac{\rho U_{max}^2a^4}{l^2}
\end{equation}

where $U_{max}$ is the maximum fluid velocity. We generalize this result to our rectangular channels (of width $W$ and height $H$) by replacing $l$ with the channel hydraulic diameter $D_h=2WH/(W+H)$. We also consider the mean flow velocity $U$ instead of the maximum flow velocity, since the former can be easily calculated based on the volumetric flow rate $Q$ and the channel cross-sectional area: $U=Q/WH$. We hence use the following scaling relationship for our rectangular channels:
\begin{equation}
    F_L =k_L \frac{\rho U^2a^4}{D_h^2}
\end{equation}

where $k_L$ is a dimensionless coefficient, and assume that it holds in curved channels as well.

The Dean’s drag acting on the particles due to the transverse Dean’s flow scales with the square of channel velocity as well (under Stokes drag) \cite{2007_DiCarlo_Toner_PNAS}:
\begin{equation}
    F_{De} \propto \frac{\rho U_{max}^2 aD_h^2}{R}
\end{equation}

Making the same substitutions in the lift force case, we thus use the following scaling for Dean’s drag in our system:
\begin{equation}
    F_{De} \propto \frac{\rho U^2 aD_h^2}{R}
\end{equation}

The scaling of the forces associated with the $K_c$ and $K_\eta$ terms of the slip velocity can be obtained by assuming Stokes drag and multiplying the velocity terms by $\eta a$, the constant of proportionality in Stokes’ law with constant coefficients of order unity removed:
\begin{eqnarray}
    F_{c} \propto \eta a^3 K_c \nabla(\dot{\gamma}\phi) \\
    F_{\eta} \propto \eta a^3\phi\dot{\gamma}K_\eta \nabla(\log \eta)
\end{eqnarray}

The particle volume fraction $\phi$ is given by the product of the volume of a single particle and its number density, hence it scales as $a^3 N$. The shear rate is a generalized velocity gradient, hence it scales as $U/D_h$, the ratio between the characteristic velocity scale and the characteristic channel length-scale. The gradient operator introduces an additional scaling factor of $1/D_h$. Thus, both $F_{c}$ and $F_{\eta}$ share a similar scaling relationship:
\begin{eqnarray}
    F_{c} \propto\ \frac{K_c\eta Ua^6N}{D_h^2} \\
    F_{\eta} \propto \frac{K_{\eta}\eta Ua^6N}{D_h^2}
\end{eqnarray}

\section{Experimental methods}

\subsection{Spiral filter fabrication}

The spiral filters were fabricated in two halves made of polycarbonate. The lower half contained spiral channels milled using a Minitech Mini-mill onto a length x width chip of thickness polycarbonate. The upper half was blank and made of thickness polycarbonate. After milling, the chips were chemically polished in an atmosphere of dichloromethane. The chip halves were then bonded under load at temperature for time.

After bonding, the spiral filters were filled with fluorescent dye (Alexa Fluor 647) and imaged in a confocal microscope, which affords a transverse view of the microchannel cross-section. The filters were imaged at four locations over the chip. Three representative filters were sawed open after confocal imaging using an electric saw for direct measurement using an optical microscope. Upon comparing the (background-subtracted) confocal intensity profile to the direct measurement results, we observe that the channel walls are located at the transverse positions where the confocal intensity value decreases to 40\% of its maximum intensity. We thus defined the channel height to be the vertical distance between the 40\% points of the confocal intensity profile.

\subsection{Cell flow protocol}
Chinese Hamster Ovary K1 cells expressing green fluorescent protein (CHO-GFP) were cultured in suspension (media based on HyClone PF-CHO multipowder system with 6 mM glutamine, 0.1\% Pluronic F-68, and 600 ug/ml G418) until a cell number density N exceeding 1e7 cells/ml at a viability exceeding 98.5\% (measured using a Beckman Coulter Vi-Cell XR). The cell suspension was concentrated to the desired number density (varying from $1\times 10^7$ cells/ml to $1.6\times 10^8$ cells/ml) by centrifugation at 1000 RPM for 8 minutes. This supernatant was used to dilute aliquots of the cell suspension as needed. The cell suspension was loaded into plastic syringes mounted in a syringe pump system (Harvard Apparatus PHD ULTRA) to be injected into the inlet of the spiral sorter chip at a constant flow rate. The spiral sorter chip was secured on an optical microscope (Nikon Eclipse Ti), illuminated with blue light, and viewed through a GFP-fluorescence imaging filter. As the cell suspension was pumped through the microfluidic chip, the optical microscope was used to capture top-down profiles of the fluorescence intensity distribution across the microchannels. 

\section{Effect of changing the flow rate and channel aspect ratio}\label{sup_sec:fr_ar}

\begin{figure}
\includegraphics[width = 3.3 in]{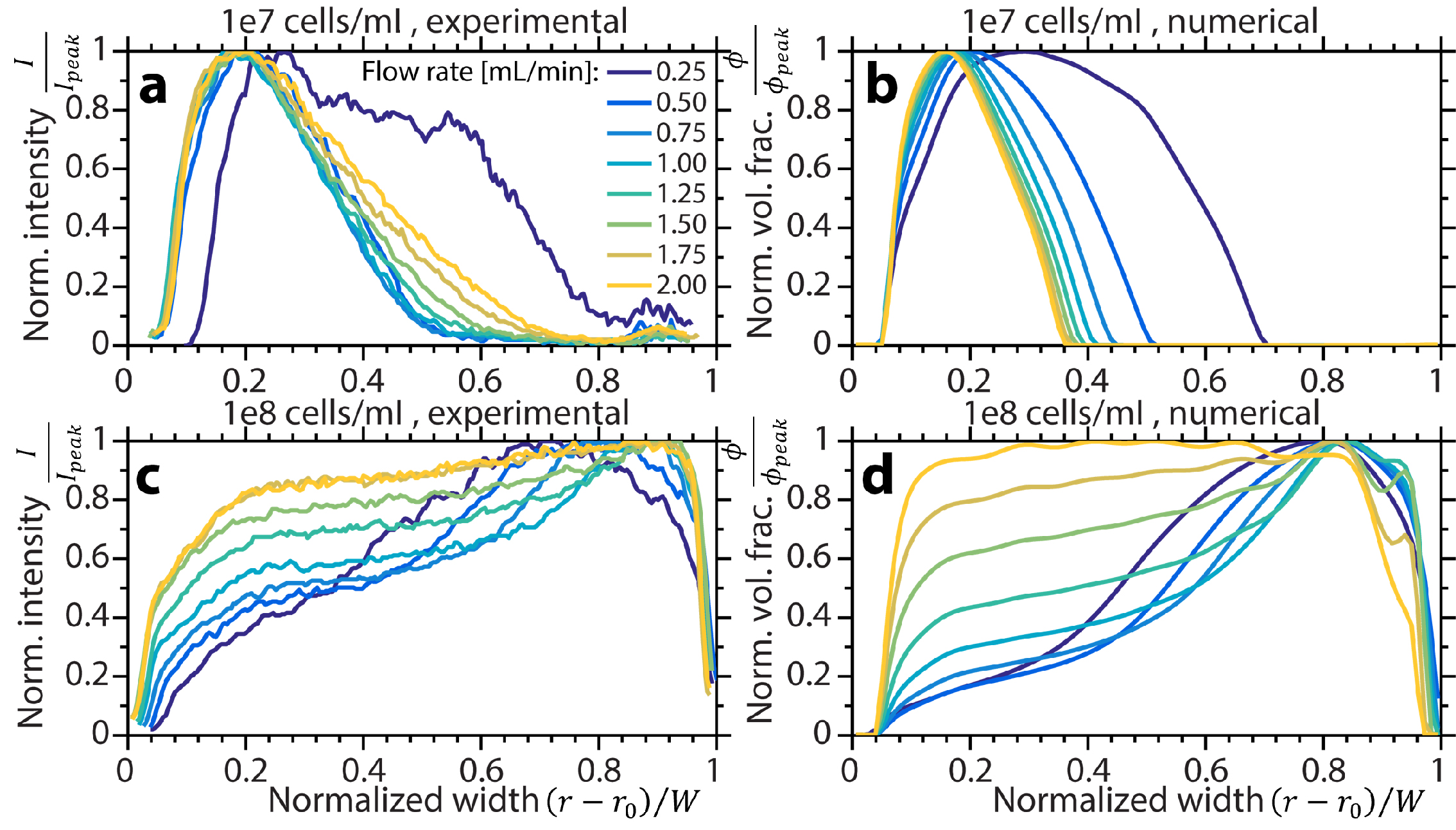} 
\caption{\label{fig:Qsweep} Dependence of focusing on flow rate for low and high cell densities and comparison with numerical predictions. The spiral channel for these experiments has dimensions $500 \mu m \times 98 \mu m$ and $R = 2.854$ mm. (a) Normalized experimental intensity profiles (normalized to the maximum intensity of each curve) for an input concentration of $10^7$ cells/ml demonstrate that the inner wall focusing profiles are insensitive to changes in the flow rate beyond 0.5 ml/min. (b) The numerical prediction for the $10^7$ cells/ml profiles using the experimental parameters in (a) (normalized to the maximum particle volume fraction in each curve) bears close resemblance to the experimental profiles. (c) Normalized experimental intensity profiles for an input concentration of $10^8$ cells/ml demonstrate better outer wall focusing for flow rates around 0.75 ml/min. Beyond $Q = 1$ ml/min, increases in flow rate cause the outer wall peak to broaden in the direction of the inner wall. (d) Numerical prediction for the profiles using the experimental parameters in (c) bear close resemblance to the experimental profiles. }
\end{figure}

We validate the numerical model by varying the volumetric flow rate under low ($10^7$ cells/ml) and high ($10^8$ cells/ml) concentrations. Figs. \ref{fig:Qsweep}(a) and (c) exhibit the fluorescence intensity profiles for initial concentrations of $10^7$ and $10^8$ cells/ml under various flow rates. Figs. \ref{fig:Qsweep}(b) and 2(d) exhibit the corresponding numerical predictions for the intensity profile under the same parameters. We obtain qualitative correspondence between the experimental and numerical profiles for flow rates between 0.25 ml/min and 2 ml/min. We observe that at $10^7$ cells/ml, the experimental system exhibits a broadening of the inner wall peak in the direction of the outer wall, whereas the numerical predictions do not produce that effect. This could be due to the experimental system entering a pre-turbulent regime at high flow rates, which introduces turbulent flow effects that we do not account for in our laminar flow numerical model. 

To avoid recalculating the lift force profile for each change of channel geometry, we propose that a single lift force profile $\vec{F}_L (r,z)$ (computed for channel dimensions $W\times H$ and hydraulic diameter $D_h$) can be extrapolated to other geometries through the lift force coefficient $\vec{k}_L$. $\vec{k}_L$ is expressed in normalized width and height coordinates using the scaling relationship adapted from Ho and Leal \cite{1974_Ho_Leal_JFM}, written here in vector form using transverse coordinates $(r,z)$:
\begin{equation}
    \vec{F}_L(r,z) = \vec{k}_L(r/W,z/H)\frac{\rho U^2a^4}{D_h^2}
\end{equation}

The velocity $U$ is given by the mean velocity $U=Q/WH$. The lift force coefficient can then be applied to other channel geometries and flow parameters:
\begin{equation}
    \vec{F}'_L(r,z) = \vec{k}_L(r/W',z/H')\frac{\rho' (U')^2(a')^4}{(D_h')^2}
\end{equation}

\begin{figure}
\includegraphics[width = 3.3 in]{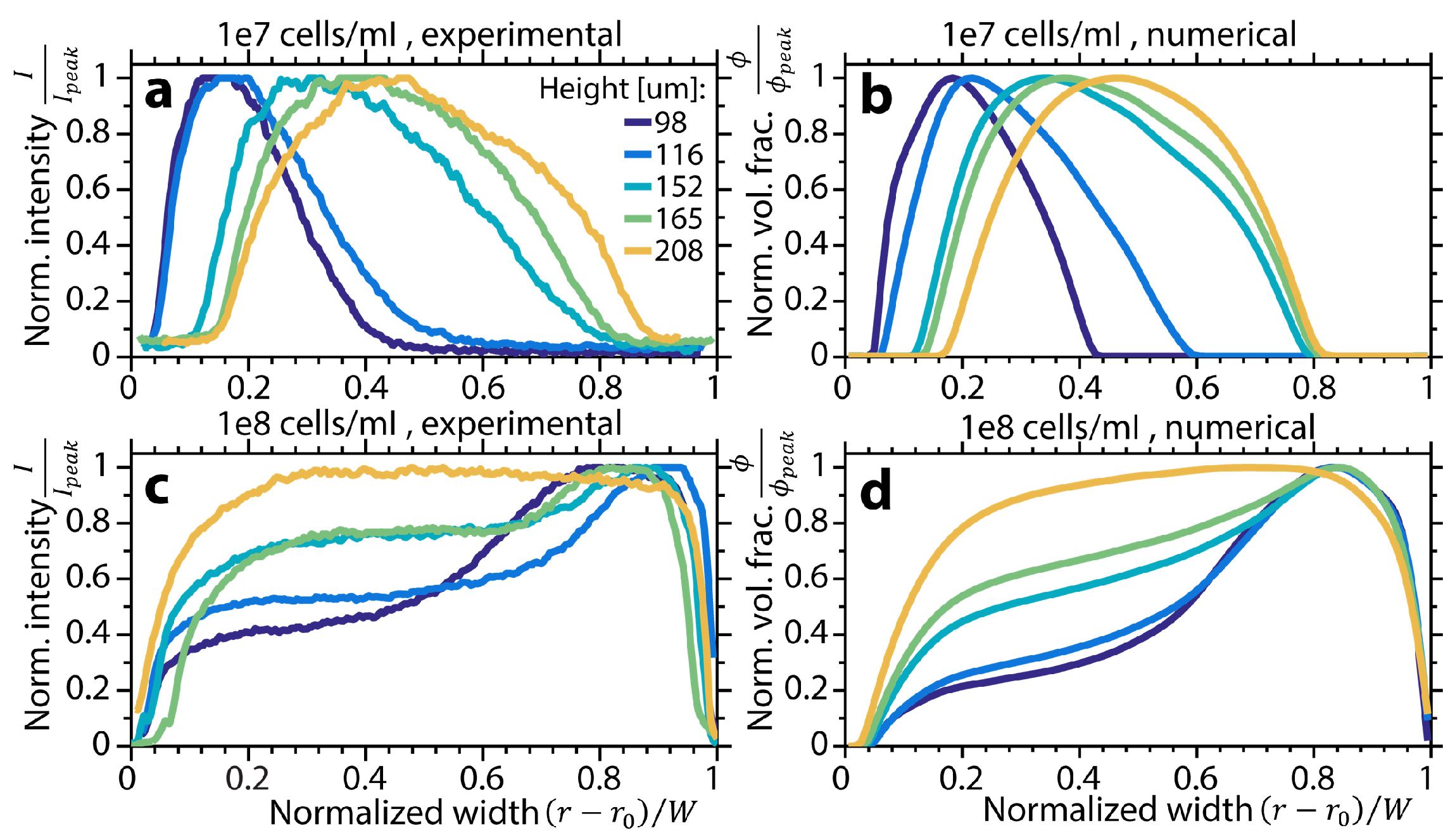} 
\caption{\label{fig:heightsweep} Dependence of focusing on channel height for low and high cell densities and comparison with numerical predictions. The spiral channel for these experiments has $W=500 \mu m, Q = 0.75$ ml/min, and $R = 2.854$ mm. (a) Normalized experimental intensity profiles (normalized to the maximum intensity of each curve) for an input concentration of $10^7$ cells/ml demonstrate well-defined inner wall focusing for shorter channels. (b) The numerical prediction for the profiles using the experimental parameters in (a) (normalized to the maximum particle volume fraction in each curve) bears close resemblance to the experimental plot. (c) Normalized experimental intensity profiles for an input concentration of $10^8$ cells/ml demonstrate superior outer wall focusing for shorter channels. (d) Numerical prediction for the profiles using the experimental parameters in (c). 
}
\end{figure}

To demonstrate the validity of the scaling approach for lift force computation across various channel aspect ratios, we fabricated spiral sorters with various channel heights and recorded the fluorescence intensity profile at low ($10^7$ cells/ml) and high ($10^8$ cells/ml) concentrations, all at $W=500 \mu m,Q=0.75 \text{ ml/min},R =2.854$ mm. The experimental profiles are plotted in Figs. \ref{fig:heightsweep}(a) and (c) for the $10^7$ cells/ml and $10^8$ cells/ml concentrations, respectively. We also used the rescaled lift force profile to predict concentration profiles for those channel heights. The numerical profile predictions are plotted in Figs. \ref{fig:heightsweep}(b) and (d). We observe close correspondence between the experimental profiles and numerical predictions for the channel heights studied, indicating that a single lift force profile can be used to predict the effect of changing channel geometries, avoiding the computationally-expensive lift force re-calculation and enabling rapid evaluation of system designs. 

\section{Q/N dependence of flow profile}

\begin{figure*}
\includegraphics[width = 5 in]{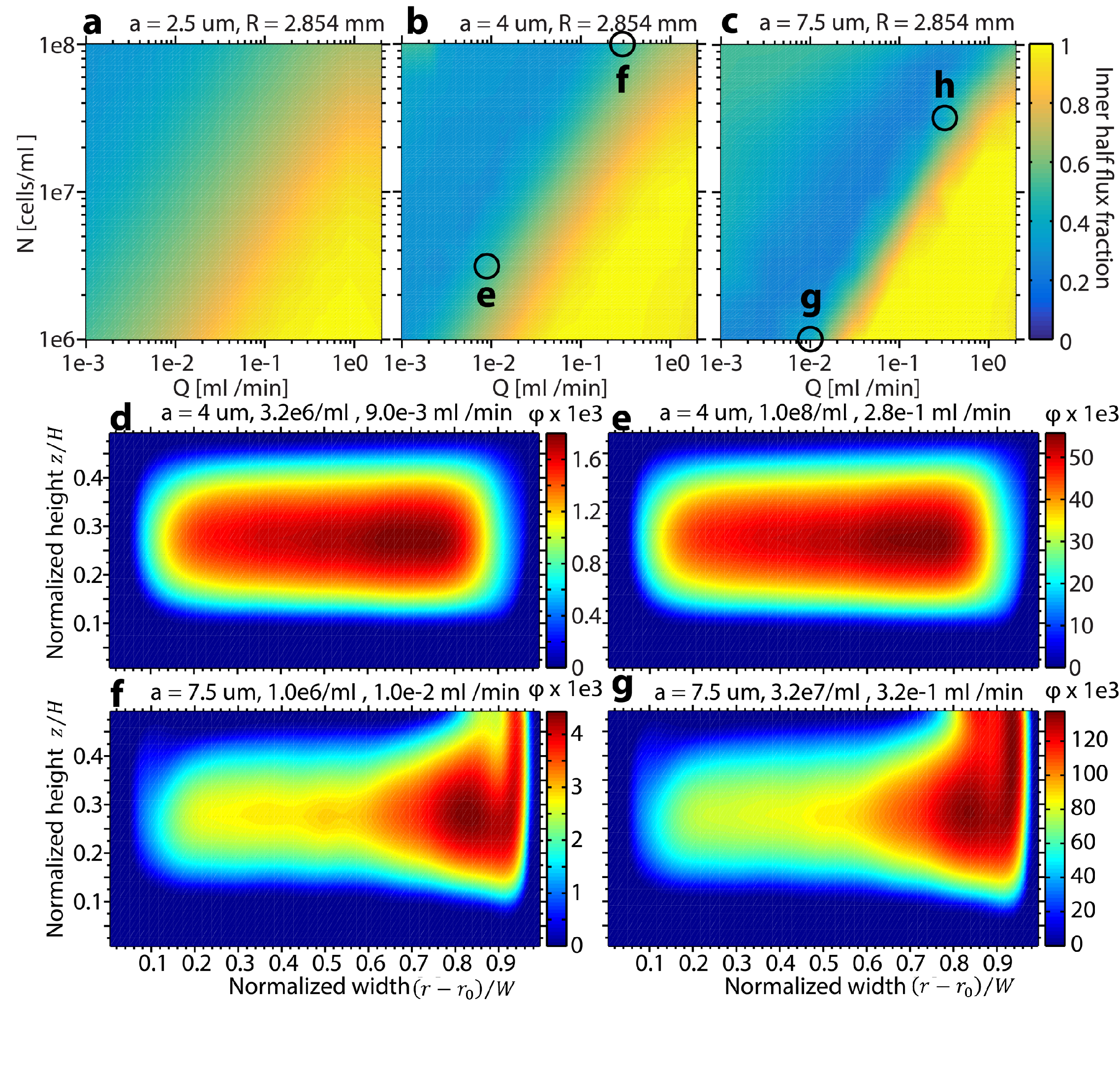} 
\caption{\label{fig:regimes} Numerical predictions for the dependence of the inner half flux fraction (fraction of particle flux located in the inner half of the channel) on volumetric flow rate Q and initial concentration N for particles similar in size to: (a) Pichia pastoris ($2.5 \mu m$ radius), (b) Red blood cells ($4 \mu m$ radius), and (c) CHO cells ($7.5 \mu m$ radius). Each case shares the same channel geometry: $W=500 \mu m,H=100 \mu m,R=2.9$ mm. Within each of the three cases, systems in which the ratio $Q/N$ is constant share the same particle distribution, hence the diagonal $45^\circ$ dependence in each surface plot. (d-g) Particle distribution comparison of pairs of points sharing the same $K=5.9\times 10^{-5} m^2/s$ for particle sizes (d-e) $a=4 \mu m$ and (f-g) $a=7.5 \mu m$. The point parameter positions are circled in the $Q-N$ plots in (b-c). Despite the pairs of points having different volumetric flow rates $Q$ and concentrations $N$, they share virtually identical particle distributions up to volume fraction scaling.}
\end{figure*}

We plot the inner half flux fraction against concentration $N$ and volumetric flow rate $Q$ for three particle radii: $2.5 \mu m$ (for \textit{Pichia pastoris} yeast, Fig. \ref{fig:regimes}(a)), $4.0 \mu m$ (for red blood cells, Fig. \ref{fig:regimes}(b)), and $7.5 \mu m$ (for CHO cells, Fig. \ref{fig:regimes}(c)). The outer wall focusing regime (low inner half flux fraction) occurs at high concentrations and low flow rates, while the inner wall focusing regime occurs at low concentrations and high flow rates. The surface plot is observed to be translationally invariant in the $45^\circ$ diagonal direction in $\log N, \log Q$ space, indicating that the inner half flux fraction depends on $Q$ and $N$ through $Q/N$. Increasing the flow rate $Q$ has the same effect as decreasing the concentration $N$. We note that this tendency is in the opposite direction as that observed by Wu \textit{et al} in trapezoidal microchannels, where outer wall focusing is associated with increases in the channel Reynolds number (which is proportional to the volumetric flow rate) \cite{2012_Wu_Han_AnalyticalChem}. We believe that this difference occurs due to the enhanced role the Deans flow plays in asymmetric channels such as the trapezoidal channel.

As examples of the similarity relationship between systems with the same value of $Q/N$, we pick two points each from the surface plots (circled and labelled on the surfaces) in Figs. \ref{fig:regimes}(b) and (c) and plot their cross-sectional volume fraction distribution. The cross-sections in Figs. \ref{fig:regimes}(d) and \ref{fig:regimes}(e) are associated with a particle radius of $4 \mu m$, and those in Figs. \ref{fig:regimes}(f) and \ref{fig:regimes}(g) are associated with a particle radius of $7.5 \mu m$. The system pairs exhibit the same particle volume fraction distribution up to overall scaling. 



\end{document}